\documentclass[12pt]{JHEP} 
\usepackage{epsfig}
\def \be {\begin{equation}}
\def \ee {\end{equation}}
\def \bea {\begin{eqnarray}}
\def \eea {\end{eqnarray}}

\def \ba {\begin{array}}
\def \ea {\end{array}}

\def \ra {\rangle}
\def \a {\alpha}
\def \b {\beta}

\def \m {\mu}
\def \n {\nu}

\title{Gravitational Radiation of Rolling Tachyon}

\author{Bin Chen\\
Department of Physics\\
KAIST\\
Daejon, 305-701\\
Korea\\
\email{chenb@muon.kaist.ac.kr} }

\author{Miao Li\\
Institute of Theoretical Physics\\
Academia Sinica\\
Beijing 100080\\
and\\
Department of Physics\\
National Taiwan University\\
Taipei 106, Taiwan\\
\email{mli@phys.ntu.edu.tw} }

\author{Feng-Li Lin\\
Physics Division\\
National Center for Theoretical Science\\
National Tsing-Hua University\\
Hsinchu 300, Taiwan\\
\email{fllin@phys.cts.nthu.edu.tw} }

\abstract{ In this paper we compute the radiation of the massless
closed string states due to the non-vanishing coupling to the
rolling open string tachyon with co-dimensions larger than 2, and
discuss the effect of back reaction to the motion of the rolling
tachyon. We find that for small string coupling, the tachyonic
matter remains in the late time, but it will completely evaporate
away over a short time of few string scales if the string coupling
is huge. Comment on the implication of our results to the
g-theorem of boundary conformal field theory is given.}

\begin{document}

\section{Introduction}

 The tachyon problem remained a significant problem in the past,
ever since the bosonic string was found to contain a tachyon in
its spectrum. It appears to be a universal feature of string
theory when supersymmetry is broken at the tree level, thus the study
of the fate of tachyon is an important problem. In the past few
years string theory has gained tremendous progress in
understanding the symmetry nature of the theory itself, along the
way, both the stable and unstable static D-branes were found.
Especially, the unstable D-brane solutions provide an arena for
studying the tachyon condensation dynamics, and many physical
issues about the endpoint of the tachyon condensation have been
clarified. In contrast, there is little understanding about the
dynamical nature of the tachyon motion itself, namely, its
time-dependent behavior and the couplings to the (time-dependent)
background spacetime.

  In analogue to the difference between the electrostatics
and electrodynamics, a novel feature of these time-varying
solutions is that they can produce the gravitational radiations
which will then react back to the motion of the tachyon and the
background geometry. Once the radiation rate of a rolling tachyon
is computed, one can plug it back to the equation of motion for
the tachyon to compute the back-reaction. We shall show in this
paper that radiation is a generic feature of a rolling tachyon,
and for simplicity we will consider a unstable D-brane embedded in
spacetime with codimension equal to or greater than 3. Other cases
require separate study. We shall show that once the string
coupling is increased, the radiation is comparable to the initial
energy stored in the unstable D-brane, and it is possible for the
D-brane to complete release its energy into the bulk if the
coupling is sufficiently large. In this case, the decay time is
the string scale. However, we will see that the numerical factors
require a large string coupling constant in order to have
significant energy loss, and the complete decay of the unstable
brane, if possible, may takes a longer time.

  We shall not consider the interesting case when the transverse
space is compactified. This problem is interesting in the
cosmological context, when one attempts to take the rolling
tachyon into account in any study of a cosmological scenario, or
to take the tachyon as an inflaton, since the reheating problem
was pointed out to be a severe problem in some tachyon-driven
inflation scenario, and the decay of tachyon may be of significant
importance in resolving this problem. Also, the case of
compactified transverse space is related to the case of D-brane
with less than 3 transverse space.

 A related problem is the S-brane, recently studied by
Gutperle and Strominger in \cite{Sbrane}. They have proposed a new
type of branes with spacelike worldvolumes which is related to the
dynamical process of the open string tachyon condensation. The
conjectured supergravity solutions of the above ``S-branes" are
constructed in \cite{Sbrane} and \cite{branesugra} based on the
isometry of the brane geometry. The S-brane configuration is
essentially different from Sen's rolling tachyon which describes
the motion of the unstable D-brane tachyon in open string (field)
theory. The difference lies in the initial conditions. In case of
S-brane, some fine tuned initial conditions are required.
Nevertheless, one can see the radiation effects in the S-brane
configuration. See \cite{Sbrane} and \cite{Strominger2} for the
related discussions and the new developments.

    Both S-brane and the rolling tachyon can be described
by the on-shell string dynamics, and the corresponding boundary
states, which encode both the open and closed string dynamics, are
constructed in \cite{Sbrane} and \cite{Sen1}. However, the
interactions between the rolling tachyon and the closed string
modes by using these boundary states has not been extensively done
yet, see the last three references in \cite{rolling} for related
consideration.

    The other interesting point in considering the gravitational
radiation is that the energy density stored in the brane can be
interpreted as the boundary entropy $g$ \cite{gtheorem}, therefore
the energy loss due to the radiation implies decrease of $g$ as
tachyon rolls, this manifest a generalized $g$-theorem in the context of
boundary conformal field theory, since coupling to closed strings
is described as loop effects in the open string picture.

  The paper is organized as following: In the next section we
first generalize Sen's rolling tachyon for the D$25$-brane to the
one with transverse directions, and calculate the couplings
between rolling tachyon boundary states and the massless closed
string modes, then obtain the linear fluctuations representing the
gravitational radiations in the asymptotic regions. In the section
3, we will calculate the radiation power accounting for the energy
carried away to asymptotic infinity by the gravitational
radiations. In section 4, we will use the energy loss rate
calculated in section 3 as the back reaction to the time evolution
of the rolling tachyon. In section 5, we will conclude with
comments and discussions, especially on the implication of the
gravitational radiation to a generalized $g$-theorem of the
boundary conformal field theory when taking into account the loop
effects in open string picture, which is the re-interpretation of
the coupling to closed string, especially to the massless mode,
i.e. gravitational radiation.

We perform all calculations in the bosonic string theory, ignoring
the existence of the closed string tachyon. Calculations in super
string theories are similar so we shall not do them here.

It remains to clarify the physical significance of the exponential
growth of massive string states in the boundary state, a
phenomenon observed in a couple of references in \cite{rolling}.

\section{Gravitational radiations from rolling tachyon}
In \cite{Sen1} Sen considered the rolling tachyon by Wick rotating
the usual marginal operator $T(X)$ of the boundary conformal field
theory(BCFT)
\be
T(X)=\lambda \cos(X) \longrightarrow T(X^0)=\lambda \cosh(X^0)\;,
\ee
which describes the rolling of the tachyon initially displaced
away from the top of the potential by $T(X^0=0)=\lambda$. After
using some tricks in BCFT, the boundary state of the rolling
tachyon up to the $(1,1)$-level in the time direction is
constructed and given as \cite{Sen1}
\be
\label{boundarystate} |B>={\cal
N}[f(X^0)+\a^0_{-1}\bar{\a}^0_{-1}g(X^0)]\delta^{d_t}(x^{\a})
\exp\{-\sum^{\infty}_{n=1}\a^{\mu\dagger}_nC_{\mu\nu}\bar{\a}^{\nu\dagger}_n\}|0\ra
\ee
with $C_{\mu\nu}=(\delta_{\a\b}, -\delta_{ij})$ and $|0\ra\equiv
|k=0\ra$ the zero-momentum state. We have generalized the rolling
tachyon on the D$25$-brane in \cite{Sen1} to the one on the
D$p$-brane. For this, the indices of the first entry of $C_{\m\n}$
run along the spatial transverse directions, and the ones in the
second entry run along the longitudinal directions. Here we assume
the spacetime dimension be $d$ and the rolling tachyon be living
on the spatial hypersurface of the D$p$-brane so that the
dimensions of the spatial transverse directions be $d_t=d-p-1$.

Computation of the gravitational radiation of a rolling tachyon in
a type II string theory is similar to what we shall do here and we
will not perform it in this paper.

  The rolling profiles are determined in \cite{Sen1} as
\be
f(x^0)={1\over 1+e^{x^0}\sin(\lambda \pi)}+{1\over
1+e^{-x^0}\sin(\lambda \pi)}-1\;, \qquad
f(x^0)+g(x^0)=\cos(2\lambda \pi)+1\;.
\ee
by treating boundary state $|B\ra$ as the source to the closed
string field $|\Phi_c\ra$ satisfying the linearized equation of
motion $(Q_B+\bar{Q}_B)|\Phi_c\ra=|B\ra$ where $Q_B$ is the BRST
operator. From the conservation equation $(Q_B+\bar{Q}_B)|B\ra=0$
one can read out the stress tensor which turns out to be
\be
\label{Tmn1} T_{00}=T_p(f(x^0)+g(x^0))\delta^{d_t}(x^{\a})\;,
\qquad T_{ij}=-2\delta_{ij}T_pf(x^0)\delta^{d_t}(x^{\a})\;,\qquad
T_{0i}=0\;,
\ee
and
\be
\label{Tmn2} T_{\a\b}=T_{\a i}=T_{0\a}=0\;,
\ee
yielding the constant energy density($\sim$ brane tension $T_p$)
and pressureless tachyonic matters in the late time
\cite{Sen1}.\footnote{In \cite{Sen1} the third equation of
(\ref{Tmn1}) is absent since there is no transverse directions
there, we will see later that this is the case if there are some.}

  At this point, it is interesting to point out that the S$(p-1)$-brane
boundary state up to the $(1,1)$-level in the time direction also
takes the form of (\ref{boundarystate}) but with the rolling
profiles
\be
f(x^0)=-g(x^0)=\delta(x^0)\;.
\ee
This implies zero energy density for the S$(p-1)$-brane.

   Since the boundary state encodes all the closed and open string
dynamics about the rolling tachyon, we can consider the coupling
between the massless closed string modes and the boundary state
(\ref{boundarystate}) to extract the gravitational perturbations
in the asymptotic flat region as in the case of the static D-brane
given in \cite{Vecchia1}. Moreover, because the source is
time-varying, one would expect it produce the gravitational
radiation once the string coupling is turned on. This radiation
will then back react to the tachyon motion and the background
geometry.

     Following \cite{Vecchia1}, the asymptotic metric and dilaton
perturbations can be read off from the current
\be
J_{MN}=<V_c|D|B>
\ee
where
\be
|V_c>=\a^M_{-1}\bar{\a}^N_{-1}|k_0,k_{\a},k_i>
\ee
is the massless closed string vertex operator, and
\be
D=\frac{1}{2\pi}\int_{|z|\leq 1}\frac{d^2
z}{|z|^2}z^{L_0-1}\bar{z}^{\bar{L}_0-1}
\ee
is the closed string disk propagator. The indices $M,N$ run from
$0$ to $d-1$.

  The straightforward calculation tells us
\be
\label{jmn} J^{MN}=\left\{\ba{ll}
 \Delta(k)\tilde{g}(k_0), \hspace{3ex} &M=N=0,\\
\Delta(k) C_{\mu\nu} \tilde{f}(k_0), \hspace{3ex} &M,N\neq 0 \ea
\right.
\ee
where $\tilde{f}(k_0), \tilde{g}(k_0)$ are the fourier
transformation of the profiles $f(x_0),g(x_0)$, and
\be
\label{green2} \Delta(k)={\cal N}({-1\over
(k_0+i\epsilon)^2-k_{\a}^2})\delta^p(k_i)\;.
\ee
is the spacetime propagator of the massless closed string modes in
k-space. Take $\epsilon>0$ for the retarded propagator. Note that
the $\delta^p(k_i)$ factor in (\ref{green2}) implies no wave
propagation along the longitudinal directions.

   Using the convention in \cite{Vecchia1} for the polarization
projection of (\ref{jmn}), we obtain the asymptotic fluctuations
of the dilaton and the metric:
\bea
{1\over
\kappa}\;\phi&=&-\Delta(k)[(1+2b)\tilde{g}(k_0)+(2p-d+3+2b)\tilde{f}(k_0)]\;,\label{phi0}\\\label{h00}
{1\over 2\kappa}\; h_{00}&=&-{\Delta(k) \over
d-2}[(3-d+2b)\tilde{g}(k_0)+(2p-d+3+2b)\tilde{f}(k_0)]\;,\\\label{htt}
{1\over 2\kappa}\;h_{\a\b}&=&{\Delta(k) \over
d-2}[(1+2b)\tilde{g}(k_0)+(2p+1+2b)\tilde{f}(k_0)]\delta_{\a\b}\;,\\\label{hll}
{1\over 2\kappa}\;h_{ij}&=&{\Delta(k) \over
d-2}[(1+2b)\tilde{g}(k_0)+(2p-2d+5+2b)\tilde{f}(k_0)]\delta_{ij}\;,
\eea
where  $b$ is an arbitrary gauge parameter due to the freedom in
choosing the polarization vector,i.e. $b=k_0l_0$ in
\cite{Vecchia1}.  These fluctuations are propagating along the
transverse directions, they are the asymptotic gravitational
radiations from the rolling tachyon.

These fluctuations can also be understood as the linear
perturbation around the flat spacetime of the following dilatonic
gravity\footnote{We use the conventions of \cite{Weinberg} for the
curvature tensor, which is related to the others' by flipping the
sign of the curvature tensor.} in the Einstein frame
\cite{Vecchia1}
\be
\label{action1} S=-{1\over 2\kappa^2}\int d^dx \sqrt{-g}[R+{1
\over d-2}(\partial \phi)^2+\cdots]
\ee
where we omit the couplings of the dilaton to the form fields.

  We can read off the source stress tensor from the linearized
Einstein equation
\be
\bar{h}_{MN,K}^{\;\;\;\;\;,K}-\bar{h}^K_{M,KN}-\bar{h}^K_{N,KM}+\eta_{MN}\bar{h}^{KL}_{,KL}
=-2\kappa^2 T_{MN}\;,
\ee
where $\bar{h}_{MN}\equiv h_{MN}-{1\over 2}\eta_{MN} h$, and from
(\ref{h00}) to (\ref{hll}) we have
\bea
{1\over 2\kappa}\;\bar{h}_{00}&=&\Delta(k)(1+b )[\tilde{g}(k_0)+
\tilde{f}(k_0)]\;,\label{hb00}\\
{1\over 2\kappa}\;\bar{h}_{\a\b}&=&- \Delta(k) b[\tilde{g}(k_0)+
\tilde{f}(k_0)]\delta_{\a\b}\;,\label{hbtt}\\\label{hbll} {1\over
2\kappa}\;\bar{h}_{ij}&=&-\Delta(k)[b(\tilde{g}(k_0)+\tilde{f}(k_0))+2\tilde{f}(k_0)]\delta_{ij}\;.
\eea
Surprisingly they are in the nice forms independent of the
dimensional parameters $p$ and $d$, moreover, all the
$b$-dependence comes with the $f+g$ which is constant in time and
therefore will not contribute to the radiations.

   In order to read off the source stress tensor, it is better to
have the Lorentz condition $k_M\bar{h}^M_N=0$, however, it is
ambiguous to see if the Lorentz condition holds because it will
involve the quantity
$k_0\Delta(k)(\tilde{f}(k_0)+\tilde{g}(k_0))\sim k_0{1\over
(k_0+i\epsilon)^2-k^2_{\a}}\delta(k_0)$ for the rolling tachyon,
which can be either zero or blow up if the mass-shell condition is
also imposed. But the pole in $\Delta(k)$ can be properly
regularized by the usual Green's function method so that we can
set it to zero and the Lorentz condition holds. The linearized
Einstein equation is then simplified to
\be
\label{Lorentzbox} \bar{h}_{MN,K}^{\;\;\;\;\;,K}=-2\kappa^2
T_{MN}\;.
\ee
It is then straightforward to get $T_{MN}$ which turns out to be
nothing but (\ref{Tmn1}) and (\ref{Tmn2}) if we set
\be
\label{tension} {\cal N}=\kappa T_p\;.
\ee
Note that we have taken the $b=0$ gauge for simplicity.

The normalization constant $\cal N$ for the static brane is given
in \cite{Vecchia1} from the factorization of the scattering
amplitude. Moreover, ${\cal N}=\sqrt{\pi}/32 \sim 10^{-2}$ for
$d=26$ if we set the string scale $2\pi l_s$ to $1$ and is
independent of $d_t$ . For only concern of the order of magnitude,
we assume the normalization for the rolling tachyon has the same
order. From now on we will set $2\pi l_s=1$ such that
$\kappa={\cal N}/T_p=2\pi g_s$ in which our ${\cal N}$ is
different from the one in \cite{Pol} by a factor of $1/2$, this is
because we have absorbed a factor of $1/2$ into ${\cal N}$ in
defining the boundary state, that is equivalent to halving the
brane tension.

  In summary: we have derived the source's stress tensor in the context
of the linearized perturbation of the dilatonic gravity by
starting from the coupling of the boundary state to the massless
closed string modes.  Moreover, these asymptotic fluctuations are
the classical gravitational radiations from the rolling tachyon.
In the next section we will apply the above first order result and
then consider the stress tensor carried by the asymptotic
fluctuations in the second order of Einstein equation.

\section{Instantaneous radiation power}
Base upon the results in the previous section, in the following we
will calculate the radiation power due to the asymptotic wave
fluctuations.
  The radiation power per transverse solid angle measured at
spatial infinity is given by
\be
\label{dpn} {dP\over d\Omega_{d_t-2}}=\lim_{r\rightarrow \infty}\;
r^{d_t-1}\hat{x}^{\a}\; t_{\a0} \;,
\ee
where $t_{MN}$ is the stress tensor for the gravitational or
dilatonic waves carried by the asymptotic perturbations, and
$r^2=x^2_{\a}$, $\hat{x}^{\a}={x^{\a} \over r}$ where index $\a$
runs for only the transverse directions since there is no wave
propagating along the longitudinal directions, i.e. $k_i=0$.

   The stress tensor for gravity wave is encoded in Einstein
equation of the second order in $h_{MN}$, that is,
\be
t^{(g)}_{MN}={1\over
2\kappa^2}[R^{(2)}_{MN}-{1\over2}\eta_{MN}\eta^{KL}R^{(2)}_{KL}]\;,
\ee
where the 2nd order Ricci tensor is given by \cite{Weinberg}
\bea
R^{(2)}_{MN}=&&-{1\over 2}
h^{KL}[h_{KL,MN}-h_{ML,NK}-h_{kN,LM}+h_{MN,LK}] \nonumber\\
&&+{1\over 4}\bar{h}^K_{L,K}[h^L_{M,N}+h^L_{N,M}-h_{MN}^{\;\;,L}]
\nonumber\\ &&-{1\over
4}[h^M_{K,L}+h_{KL}^{\;\;,M}-h^M_{L,K}][h^{K,L}_M+h^{KL}_{\;\;,M}-h^{L,K}_M]\;.
\eea
Note that the above second line vanishes if $h_{MN}$ satisfies the
Lorentz gauge condition. In the following we will stick to this
gauge by setting $b=0$. The restriction to the Lorentz gauge will
not affect the physical result since the gauge parameter $b$ is
always associated with $f(x^0)+g(x^0)$ which is constant in time
and will not contribute to the radiation power.

    Even without knowing the explicit coordinate-space form of
$\bar{h}_{MN}$ of (\ref{h00})-(\ref{hll}) but just the fact that
$\bar{h}_{00}$ is constant in time, after some lengthy
calculation, we can obtain the following simple result
\be
\label{instant1} 2 \kappa^2 t_{0\a}= R^{(2)}_{0\a}=2c_1
\bar{h}^{(l)}_{\;,0\a}\bar{h}^{(l)}+c_1\bar{h}^{(l)}_{\;,0}\bar{h}^{(l)}_{\;,\a}+c_2
\bar{h}^{(l)}_{\;,0}\bar{h}^{(0)}_{\;,\a}\;,
\ee
where
\be
\bar{h}^{(l)}\equiv {1\over p}\; \delta^{ij}\bar{h}_{ij}\;,
\ee
and the constants $c_1$ and $c_2$ are
\bea
c_1&=&{-1\over 4}{p(d-p-2) \over d-2}={-1\over
4}{(d_t-1)(d-d_t-1)\over d-2}\;,\\
c_2&=& {1\over 4}{p\over d-2}={1\over 4}{d-d_t-1 \over d-2}\;.
\eea

  Moreover, if $\bar{h}^{(l)}$ decays exponentially in time, which is
the case for Sen's rolling tachyon profile, then we  can drop the
total derivative term with respect to time and get the simpler
result
\be
\label{grpower} 2\kappa^2 t_{0\a}=-c_1
\bar{h}^{(l)}_{\;,\a}\bar{h}^{(l)}_{\;,0}\;.
\ee

   Up to now we are considering the gravitational radiations from
the homogeneous time-varying source. As known in the 4-dimensional
electrodynamics and gravity, relating to their tensor structures,
there are only dipole electromagnetic radiation and quadrupole
gravity wave. Naively we would not expect gravity wave for a
homogeneous source, however, what we have considered is the the
gravitational waves from the extended objects in the dilaton
gravity in d-dimensional spacetime, therefore there is always the
dilaton radiation which requires no multipole structure, also
different dimensionality may require different multipole structure
for the gravity wave.  From our formula (\ref{grpower}) we see
that $c_1=0$ for $d=4$ and $d_t=3$ which agrees with the result in
4-dimensional spacetime. For the other cases there will be nonzero
asymptotic gravity radiation power as we will see in the
following.

    In order to get the explicit form for $\bar{h}^{(l)}$ in the
radiation power formula, one needs the coordinate representation
of the retarded Green's function in the asymptotic region of the
$d_t+1$ spacetime. This is not seen in the standard textbook for
general $d_t\ne 3$, so we derive it below.

   Let's start with the k-space representation of the Green's function
for $n+1$-dimensional spacetime
\be
G(\vec{x},t)=\int {d^nk dk_0 \over (2\pi)^{n+1}}{e^{i\vec{k}\cdot
\vec{x}-ik_0 t} \over (k_0+i\epsilon)^2-k^2}\;,
\ee
where $k=|\vec{k}|$.

After integrating out the $k_0$ we get
\be
\int {d^nk \over (2\pi)^n 2k} \; e^{i\vec{k}\cdot \vec{x}-ik t}\;,
\ee
and choose the spherical polar coordinates such that
\be
d^nk=k^{n-1}\sin^{n-2}\theta d\theta d\Omega_{n-2}\;,
\ee
and
\be
\vec{k}\cdot \vec{x}=kr \cos\theta\;.
\ee

Then the Green's function becomes
\be
\label{gr1} G(\vec{x},t)={\Omega_{n-2}\over2(2\pi)^n} \int
k^{n-2}I(kr,n)e^{-ikt} dk\;,
\ee
where
\be
\label{be1} I(kr,n)=\int^{\pi}_0 e^{ikr\cos\theta}\sin^{n-2}\theta
d\theta=\sqrt{\pi}\; 2^{{n-2\over 2}} \Gamma({n-1\over2}){1\over
(kr)^{{n-2\over 2}}}J_{{n-2\over 2}}(kr)\;,
\ee
In the above we have used the definition of the integral
representation of the Bessel function $J_{\nu}(z)$.

   Since we are only interested in the large $r$ behavior of the
Green's function, we can use the large $z=kr$ expansion of
$J_{\nu}(z)$, namely,
\be
J_{\nu}(z)=\sqrt{2 \over \pi z}\;\{\cos(z-{\nu\pi \over
2}-{\pi\over 4})[1+{\cal O}(1/z^2)]+\sin(z-{\nu\pi \over
2}-{\pi\over 4}){\cal O}(1/z)\}\;.
\ee
Keep the leading term in the above, which turns out to be the only
term contributing to the radiation power in the $r\rightarrow
\infty$ limit, and plug it into (\ref{be1}) and (\ref{gr1}) we get
\be
G(\vec{x},t)\sim {\Omega_{n-2}\over (2\pi)^n} \sqrt{\pi}\;
2^{{n-3\over 2}} \Gamma({n-1\over2}) r^{{1-n \over
2}}\int^{\infty}_0 dk k^{{n-3 \over 2}} e^{ik(r-t)}\;,
\ee
where we have dropped the overall constant phase and also the
advanced part proportional to $e^{-ik(r+t)}$.

  From the fact
\be
4\pi \int^{\infty}_0 dk e^{ik(r-t)}=\delta(r-t)
\ee
so that we have
\be
4 \pi\int_0^{\infty} dk k^m e^{ik(r-t)}=(i)^m{d^m \delta(t-r)
\over dt^m}\equiv (i)^m \delta^{[m]}(t-r)\;.
\ee

For $n\ge 3$ and odd,
\be
\label{greens} G(\vec{x},t)=g_n  r^{{1-n \over
2}}\delta^{[{n-3\over2}]}(t-r)\;,
\ee
where $|g_n|={\Omega_{n-2}\over \sqrt{\pi}(2\pi)^n} \;
2^{{n-7\over 2}}
\Gamma({n-1\over2})=2^{-({n+5\over2})}\pi^{-({n+2\over2})}$.

In comparison with the leading term in (\ref{greens}) which will
yield finite radiation power, the higher terms of $1/(kr)$ in
$J_{\nu}(kr)$ have more negative power $r$-dependence, therefore
these terms do not contribute to the radiation power measured at
spatial infinity.

For $n>3$ but even, we need to evaluate
\be
\int_0^{\infty}  k^{m+{1\over2}} e^{ik(r-t)} dk\;,
\ee
which has no simple form in the coordinate space. Nevertheless, it
is a simple matter to compute the total radiation energy in
k-space.

For $0<n<3$ the brane curves the transverse directions, so the
asymptotic consideration will break down, one may follow the
treatment in \cite{PW95} for the cases with one or two transverse
directions.  For $n=0$ there is no transverse directions, instead
one can follow the consideration of the tachyon cosmology
\cite{tachyonic,reheat} where the rolling tachyon matters act as
the source to the evolution of the background geometry.

  For concreteness, we consider the cases with $d_t\ge 3$ and odd in
the following. Using the Green's function (\ref{greens}) we will
get the leading term for the asymptotic fluctuations
\be
\bar{h}^{(l)}(\vec{x},t)= 4\kappa^2 T_p g_{d_t} \;{1\over
r^{{d_t-1\over 2}}} f^{[{d_t-3 \over 2}]} (t-r)\;, \qquad\;.
\ee
Combining this with (\ref{grpower}) and (\ref{dpn}) we get the
corresponding radiation power for gravity wave
\be
P_g=8c_1 \Omega_{d_t-2}[\kappa T_p \;|g_{d_t}| \; f^{[{d_t-1 \over
2}]} (t-r) ]^2.
\ee
This is the energy loss rate measured at the spatial infinity, for
the instantaneous one measured at the source we should take care
of the retarded effect and replace the argument $t-r$ by $t-r+r=t$
in the above formula.

   Similarly, we can calculate the instantaneous radiation power
due to the asymptotic dilaton wave whose stress tensor is the
standard one derived from action (\ref{action1}), that is
\be
t^{(\phi)}_{MN}\equiv -2{\delta L^{(\phi)}\over \delta
g^{MN}}+g_{MN}L^{(\phi)}={1\over 2\kappa^2}{1\over
d-2}[2\partial_M\phi \partial_N\phi-\eta_{MN} (\partial
\phi)^2]\;.
\ee

  From (\ref{phi0}) and (\ref{greens}) the dilaton fluctuation
in the coordinate space is given by
\be
\phi(\vec{x},t)= \kappa^2 T_p (d-2d_t)\;{1\over r^{{d_t-1 \over
2}}} f^{[{d_t-3 \over 2}]} (t-r)\;.
\ee
Using this we can get the instantaneous radiation power of
dilatonic wave as
\be
P_{\phi}=-{(d-2d_t)^2\over d-2} \Omega_{d_t-2}[\kappa T_p
\;|g_{d_t}| \; f^{[{d_t-1 \over 2}]} (t) ]^2.
\ee

   The total gravitational instantaneous radiation power is
\be
\label{Ptot} P=P_{g}+P_{\phi}=-c_{d,d_t}[\kappa T_p \; f^{[{d_t-1
\over 2}]} (t) ]^2\;,
\ee
where the overall constant
\be
c_{d,d_t}={2(d_t-1)(d-d_t-1)+(d-2d_t)^2 \over (d-2) ({d_t-3\over
2})! 2^{d_t+4}(\sqrt{\pi})^{d_t+5}}\;.
\ee
Note that $P$ is the radiation power per unit world-volume, we
then check that its dimension is $L^{-p-2}$ as expected. Moreover,
for any physical parameters $d$ and $d_t$, $P$ is always
negative,i.e. energy loss.

\section{Back reaction on rolling tachyon}
The energy loss due to the gravitational radiations will have the
back reaction on the tachyon motion so that the energy of the
rolling tachyon will be no longer constant in time.

  We can see from (\ref{Ptot}) that the energy loss is
independent of the string or gravitational coupling since $\kappa
T_p={\cal N}$. This justifies our far field approximation.
However, the ratio of energy loss to the origin energy density
stored in the brane is proportional to $g_s$,i.e. $\kappa={\cal
N}/T_p=2\pi g_s$. This implies that the energy loss will be
comparable to $T_p$ if $g_s$ is large enough, and be negligible if
$g_s$ is very small. We can estimate the order of $g_s$ in order
for this to be the case. By integrating $P$ over all time we get
ratio of the total energy density loss to $T_p$
\be
\label{ratio1}
{\Delta \rho \over T_p}= -2\pi g_s {\cal N}\;{c_{d,3}\over
6}[a^6+9a^4-9a^2-1-12(a^4+a^2)\ln a] (a^2-1)^{-3} \ee
for $d_t=3$;
and
\bea
\label{ratio2}
{\Delta \rho \over T_p}=&&-2 \pi g_s {\cal N} \;{c_{d,5}
\over 30} [a^{10}-125a^8-350a^6+350a^4+125a^2-1
\nonumber\\
&&+60(a^8+11a^6+11a^4+a^2)\ln a](a^2-1)^{-5}
\eea
for $d_t=5$. The parameter $a$ is the $\sin(\lambda \pi)$ in the
definition of $f(t)$.

   Both expressions for the energy loss are finite as long as
$0<a=\sin(\lambda \pi)<1$, and the explicit value for ${\cal N}
c_{d,d_t}$ shows that $g_s \sim 10^5$ for $d=26,d_t=3$ and $g_s
\sim 10^7$ for $d=26,d_t=5$ in order for the energy loss to be
comparable with the brane tension. Moreover, numerical plot of $P$
shows that it has ${d_t-1 \over 2}$ local maxima peaked around the
order of string scale, and the distribution of $P$ is more close
to $t=0$ than that for the rolling profile $f(t)$, see Figure 1
for details. This implies that it is possible for the rolling
tachyon to completely turn into the gravitational radiation if the
string coupling is large enough, resulting in no remaining
tachyonic matters in the late time. Of course, the energy loss
should not be greater than the original energy stored in the brane
to not violate the positive energy theorem in general relativity
\cite{yau}.

\begin{figure}[ht]
\center{\epsfig{figure=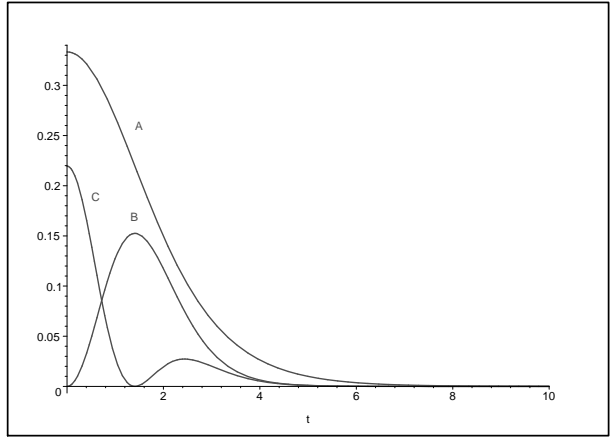,angle=-90,width=8cm}}
   \caption{This plot shows that the function $\dot{f}(t)$(the curve B) for $d_t=3$
and $\ddot{f}(t)$(the curve C) for $d_t=5$ are localized in the
early stage of the tachyon motion in contrast to the rolling
profile $f(t)$(the curve A). In the plot $\sin(\lambda \pi)$ is
set to $0.5$ and both $\dot{f}$ and $\ddot{f}$ have been scaled up
by 10 times. This indicates that the gravitational radiation
occurs in the early stage and lasts only for few stringy scale
duration.}
   \label{fig}
\end{figure}

It is interesting to study the behavior of the above ratios in the
$a\rightarrow 1$ limit\footnote{We would like to thank the remarks
of the JHEP's referee on this point.}. Note that $a=1$ or
$\lambda=1/2$ represents the closed string vacuum, and the initial
energy density stored in the tachyon is $T_p (1+cos(2\lambda
\pi))=2T_p(1-a^2)$, which is vanishing in the $\lambda \rightarrow
1/2$ limit. In this limit we should expect a vanishing value for
the above ratios (\ref{ratio1}) and (\ref{ratio2}) in accordance
with the linear response theory. However, a naive inspection of
(\ref{ratio1}) and (\ref{ratio2}) suggests that they diverge as
$(a^2-1)^{-3}$ and $(a^2-1)^{-5}$ respectively. This is actually
not true. After a careful treatment of the Taylor expansion of
$\ln(a)$ in (\ref{ratio1}) and (\ref{ratio2}) up to the 5th and
7th order respectively, we find that both ratios vanish as
$(a^2-1)^2$ and the corresponding ratio of the total energy loss
to the original energy density vanishes actually as $(a^2-1)$
which is consistent with the expectation of the linear response
theory. This gives a nontrivial consistent check of our
computation.

  We should emphasize that the validity of our far field
approximation should be independent of the magnitude of the
gravitational coupling. This is in analogue to the case of black
hole with a macroscopic huge mass, there we can trust our far
field approximation for any mass. The only difference is that our
far fields are not static. However, if $g_s$ is large enough, then
the radiation effect dominates the classical tachyon motion, and
it is hard to describe the exact motion of the rolling tachyon
even at late time. On the other hand, if $g_s$ is small, then we
can treat the gravitational radiation as a small external
"frictional" force to the tachyon motion, then we can try to see
the modified rolling behavior at the late time, namely, large time
in comparison with the string scale.

   In order to estimate the power loss rate in the late time, we
can approximate the profile $f(t)$ by keeping the leading term in
$e^{-t}$ and get
\be
f(t)\sim \cot(\pi \lambda)\cos(\pi \lambda) e^{-t}\;.
\ee
  From this we know that the total radiation power in the late time
behaves as
\be
P \sim e^{-2t}\;,
\ee
which is negligible for large $t$. Therefore, for small $g_s$ the
overall physical picture on the back reaction of tachyon motion is
that the energy of the tachyon decreases gradually in mild rate at
the early stage and then comes to a constant again in the late
time.

  One can make the statement in more quantitative sense by
invoking the effective field theory analysis of the rolling
tachyon \cite{Sen1}, which is described by the following DBI-type
Lagrangian \cite{Garousi}
\be
\label{DBI} -V(T)\sqrt{-det(\eta+\partial T \partial
T)}=-V(T)\sqrt{1-\dot{T}^2}\;,
\ee
where the tachyon potential $V(T)\sim e^{-T/2}$.

  From (\ref{DBI}) we can derive the energy density for
the system, which is
\be
\rho={V(T)\over \sqrt{1-\dot{T}^2}}\;.
\ee
If we neglect the gravitational radiations, $\rho$ is conserved,
then in the late time, the tachyon will have the solution in the
following form
\be
\label{leadingT} T=t+C e^{-t}+ {\cal O}(e^{-2t})
\ee
so that $T\rightarrow \infty$ and $\dot{T} \rightarrow 1$. This
field configuration result in the pressureless tachyonic matter in
the late time.

   By taking into account the small but non-vanishing gravitational
radiation power, $\rho$ is no longer conserved, instead
\be
{d\rho\over dt}=P\;.
\ee
Recall that $P\sim e^{-2t}$, it will not affect the leading term
behavior of $T$ in (\ref{leadingT}).

  We can also consider the loss of the momentum density $p_M$
due to the gravitational radiations  by following the same
procedure for the energy loss\footnote{We thank C.-S. Chu for the
discussion on this point.}, namely,
\be
{d p_M \over dt\;d\Omega_{d_t-2}}=\lim_{r\rightarrow \infty}\;
r^{d_t-1}\hat{x}^{\a}\; t_{\a M} \;.
\ee
It is easy to see that there is no longitudinal momentum loss and
the transverse momentum loss (recoil effect) has the same
suppression factor $e^{-2t}$ as for the energy loss in the late
time with small string coupling. Therefore, the rolling tachyon
will still form the pressureless matters in the late time,
although the energy and transverse momentum stored in the rolling
tachyon(or brane) is not conserved.

  For sufficiently large string coupling constant or
in the radiation active early stage, one needs to solve the
equation of motion in a fully consistent way by taking into
account the back reaction to the geometry, in order to have an
exact profile. This is beyond the reach of this work. Moreover,
since $g_s$ is huge the quantum gravity effect could be quite
relevant and the classical radiation would be less dominant.
However, it is interesting to notice that the recoil effect is
also huge, reflecting the necessity of studying the back reaction
to the bulk geometry for highly strong coupling case even at the
classical level.

\section{Conclusion}

   In this paper, we have calculated the gravitational radiation power
of the rolling tachyon, and consider its back reaction to the
tachyon motion. We see that for Sen's smooth rolling profile, the
perturbation calculation can be put into a nice framework of
boundary state even though the background is time-dependent.

The total radiation energy, compared to energy stored initially in
the unstable brane, is of order $g_s$. Thus naively one expects
that when $g_s$ is of order 1, the unstable brane completely decay
into massless string modes, and the time it takes is the string
scale. Surprisingly, this is not the case, since all numerical
factors compromise to require a rather large coupling constant.
When the back reaction is taken into account, the radiation can be
larger, since the tachyon profile may change faster. It requires
to solve the back reaction equation in order to see what really
happens.

It is an interesting  problem to calculate radiation when the
codimension of the unstable brane is less than 3. In case of no
codimension, we are dealing with a cosmological setting, all
possible ``energy release" takes the form of deforming the
background metric in an uniform way. Besides, due to the form of
the tachyon potential, it is difficult to get reheating in the
rolling-tachyon driven inflation scenario \cite{reheat}. If there
is no transverse direction, the reheating problem remains.
Otherwise, we would expect the energy release to the bulk if the
compactified transverse space is sufficiently large, and the back
reaction will modify the inflaton potential to solve the reheating
problem. However, it requires more detailed studies to have a
definite answer.

An implicit assumption for our calculation is that the rolling
profile is non-singular. This is in contrast to the "rolling"
profile of the S$p$-brane where $f(t)=\delta(t)$, and it yields a
singular instantaneous radiation power being proportional to
$(\delta^{[{d_t-1\over 2}]}(t))^2$, implying that the perturbation
framework breaks down.  This is not surprising since the
S$p$-brane  is ``created" and ``annihilated" in a single moment,
the energy profile is infinite which should excite all the stringy
modes and put the string theory in the Hagedorn phase where both
the quantum and thermal effects are essential. Some report in
progress along line can be found in the recent paper by Strominger
\cite{Strominger2}.

  The final point we would like to mention is that our result
implies a connection of the closed string radiations to the
$g$-theorem of the BCFT. As pointed out in \cite{gtheorem} the
tension of the unstable brane can be interpreted as the boundary
entropy $g$ which counts the dimensions of the Hilbert space of
BCFT. If one turns off the string coupling, there is no
gravitational radiation and the energy density stored in the brane
remains constant, yields no RG flow as the tachyon rolls. Once the
string coupling is turned on, we see that the energy density of
the brane is decreasing as tachyon rolls, this is a signature that
the boundary entropy $g$ is decreasing even we are only turning on
the marginal boundary deformation. However this effect comes from
interaction with closed strings, manifests itself as loop effects
in the open string picture, thus does not contradicts Sen's
result. The loop effects can be summarized at the open string tree
level (the disk BCFT), and implies that the bulk radiation plays
the role of the relevant deformation for BCFT, and it
characterizes the $g$-theorem. Moreover, the positive energy
theorem of general relativity, especially the one for the Bondi
mass \cite{yau} guarantee that $g$ is always positive as one
should expect from its definition. However, it remains a task to
give a more precise map between the Bondi mass and the quantity in
BCFT. On the other hand, the massive closed string modes will not
result in the IR divergence in the loop effect of the open string,
in accordance with the fact that they cannot carry energy flux to
infinity, therefore, they are not the relevant deformation of
BCFT.

\bigskip

\acknowledgments We appreciate the participation of Chiang-Mei
Chen and John Wang at the early stage of this project, especially
for C.M.'s many useful discussions and comments. BC and FLL would
like to thank ITP, Beijing for her hospitality during their visit
this summer. The work of BC was supported by Korea Science and
Engineering Foundation (KOSEF) Grant R02-2002-000-00146-0.



\begin{thebibliography}{99}

\bibitem{Sbrane} M. Gutperle, A. Strominger, ``Spacelike Branes",
hep-th/0202210, {\em JHEP \bf 0204} (2002) 018.

\bibitem{branesugra}
C.-M. Chen, D. V. Gal'tsov, M. Gutperle, ``S-brane Solutions in
Supergravity Theories", hep-th/0204071, {\em Phys.\ Rev.\ \bf D66}
(2002) 024043.
\\M. Kruczenski, R. C. Myers, A. W. Peet ``Supergravity S-Branes",
hep-th/0204144, {\em JHEP \bf 0205} (2002) 039.
\\S. Roy, ``On supergravity solutions of space-like Dp-branes",
hep-th/0205198.

\bibitem{Sen1}
A. Sen, ``Rolling Tachyon", hep-th/0203211, {\em JHEP \bf 0204}
(2002) 048. \\``Tachyon Matter", hep-th/0203265. \\``Field Theory
of Tachyon Matter", hep-th/0204143. \\``Time Evolution in Open
String Theory", hep-th/0207105.

\bibitem{rolling}
S. Sugimoto, S. Terashima, ``Tachyon Matter in Boundary String
Field Theory", hep-th/0205085, {\em JHEP \bf 0207} (2002) 025.
\\J. A. Minahan, ``Rolling the tachyon in super BSFT", hep-th/0205098,
{\em JHEP \bf 0207} (2002) 030.
\\
A. Ishida, S. Uehara, ``Gauge Fields on Tachyon Matter",
hep-th/0206102.
\\J.E. Wang, ``Spacelike and Time Dependent Branes from DBI",
hep-th/0207089.
\\ N. Moeller, B. Zwiebach, ``Dynamics with Infinitely Many Time
Derivatives and Rolling Tachyons", hep-th/0207107.
\\ P. Mukhopadhyay, A. Sen, ``Decay of Unstable D-branes with
Electric Field", hep-th/0208142.
\\T. Okuda, S. Sugimoto, ``Coupling of Rolling Tachyon to Closed
Strings", hep-th/0208196.
\\N.D. Lambert, I. Sachs, ``Tachyon Dynamics and the Effective Action
Approximation", hep-th/0208217.
\\ G. Gibbons, K. Hashimoto, P. Yi, ``Tachyon Condensates, Carrollian
Contraction of Lorentz Group, and Fundamental Strings",
hep-th/0209034.


\bibitem{tachyonic} G.W. Gibbons, ``Cosmological Evolution of the
Rolling Tachyon", hep-th/0204008, {\em Phys.\ Lett.\ \bf B537}
(2002) 1-4.
\\
S. Mukohyama, ``Brane cosmology driven by the rolling tachyon", ,
hep-th/0204084, {\em Phys.\ Rev.\ \bf D66} (2002) 024009.
\\
T. Padmanabhan, ``Accelerated expansion of the universe driven by
tachyonic matter", hep-th/0204150, {\em Phys.\ Rev.\ \bf D66}
(2002) 021301.
\\
D. Choudhury, D. Ghoshal, D. P. Jatkar, S. Panda, ``On the
Cosmological Relevance of the Tachyon", hep-th/0204204.
\\
X.-Z. Li, J.-G. Hao, D.-J. Liu, ``Can Quintessence Be The Rolling
Tachyon?", hep-th/0204252, {\em Chin.\ Phys.\ Lett.\ \bf 19}
(2002)1584.
\\
G. Shiu, I. Wasserman, ``Cosmological Constraints on Tachyon
Matter", hep-th/0205003, {\em  Phys.\ Lett.\ \bf B541} (2002)
6-15.
\\
T. Mehen, B. Wecht, ``Gauge Fields and Scalars in Rolling Tachyon
Backgrounds", hep-th/0206212.
\\
K. Ohta, T. Yokono, ``Gravitational Approach to Tachyon Matter",
hep-th/0207004.
\\
G. Shiu, S.-H. Tye, I. Wasserman, ``Rolling Tachyon in Brane World
Cosmology from Superstring Field Theory", hep-th/0207119.
\\
A. Buchel, P. Langfelder, J. Walcher, ``Does the Tachyon Matter?",
hep-th/0207235.
\\
G. Felder, L. Kofman, A. Starobinsky, ``Caustics in Tachyon Matter
and Other Born-Infeld Scalars", hep-th/0208019.
\\
S. Mukohyama, ``Inhomogeneous tachyon decay, light-cone structure
and D-brane network problem in tachyon cosmology", hep-th/0208094.
\\
J.-G. Hao, X.-Z. Li, ``Reconstructing the Equation of State of
Tachyon", hep-th/0209041.
\\
A. Sen, ``Time and Tachyon", hep-th/0209122.


\bibitem{Vecchia1} P. Di Vecchia, M. Frau, I. Pesando, S. Sciuto, A.
Lerda, R. Russo, ``Classical p-branes from boundary state",
hep-th/9707068, {\em Nucl.\ Phys. \bf B507} (1997) 259-276.

\bibitem{Garousi} M. R. Garousi, ``Tachyon couplings on non-BPS D-branes and Dirac-Born-Infeld
action", hep-th/0003122, {\em Nucl.\ Phys.\ \bf B584} (2000) 284.
\\
 E. A. Bergshoeff, M. de Roo, T. C. de Wit, E. Eyras, S. Panda, ``T-duality and Actions for Non-BPS
 D-branes", hep-th/0003221, {\em JHEP \bf 0005} (2000) 009.


\bibitem{Weinberg}S. Weinberg, ``Gravitation and Cosmology: Principles
and Applications of The General Theory of Relativity", John Wiley,
1972.

\bibitem{Pol} J. Polchinski, ``TASI Lectures on D-Branes",
hep-th/9611050.

\bibitem{PW95} J. Polchinski, E. Witten, ``Evidence for Heterotic -
Type I String Duality", hep-th/9510169, {\em Nucl.\ Phys.\ \bf
B460} (1996) 525-540.


\bibitem{reheat} M. Fairbairn, M. H.G. Tytgat, ``Inflation from a Tachyon Fluid?",
hep-th/0204070.
\\
A. Feinstein, ``Power-Law Inflation from the Rolling Tachyon",
hep-th/0204140, {\em Phys.\ Rev.\ \bf D66} (2002) 063511.
\\
L. Kofman, A. Linde, ``Problems with Tachyon
Inflation", hep-th/0205121, {\em JHEP \bf 0207} (2002) 004.
\\
Y.-S. Piao, R.-G. Cai, X.-M. Zhang, Y.-Z. Zhang, ``Assisted
Tachyonic Inflation", hep-ph/0207143.
\\
J. M. Cline, H. Firouzjahi, P. Martineau, ``Reheating from Tachyon
Condensation", hep-th/0207156.
\\
 M. C. Bento, O. Bertolami, A.A. Sen, ``Tachyonic Inflation in the Braneworld
 Scenario", hep-th/0208124.

\bibitem{yau} R. Schoen, S.-T. Yau, ``On the Proof of the Positive Mass Conjecture in General
Relativity", {\em Commun.\ Math.\ Phys.\ \bf65} (1979)45-76;
``Positivity of the Total Mass of a General Space-Time", {\em
Phys.\ Rev.\ Lett.\ \bf 43} (1979), 1457; ``Proof that the Bondi
Mass Is Positive", {\em Phys.\ Rev.\ Lett.\ \bf 48} (1982), 369.



\bibitem{gtheorem} J. A. Harvey, D. Kutasov, E. J. Martinec, ``On the Relevance of
Tachyons", hep-th/0003101.
\\J. A. Harvey, S. Kachru, G. Moore, E. Silverstein, ``Tension is
Dimension", hep-th/9909072, {\em  JHEP \bf 0003} (2000) 001.

\bibitem{Strominger2} A. Strominger, ``Open String Creation by
S-branes", hep-th/0209090.

\end{thebibliography}
\end{document}